\documentclass[12pt]{article}
\usepackage[margin=0.95 in]{geometry}
\usepackage{amsmath}
\usepackage{amssymb,amsfonts}
\usepackage[all]{xy}
\usepackage{graphicx}
\usepackage[utf8x]{inputenc}
\usepackage{amsmath}
\usepackage{amssymb}
\usepackage{float}
\usepackage{array}
\usepackage{tikz}
\usepackage{mathtools}
\usepackage{mathrsfs} 
\usepackage{hyperref}  
\usepackage{cite}
\numberwithin{equation}{section}
\numberwithin{equation}{section}
\numberwithin{table}{section}
\begin{document}

\thispagestyle{empty}

\vspace*{3cm}
{}

\noindent
{\LARGE \bf  An Elliptic Triptych }
\vskip .4cm
\noindent
\linethickness{.06cm}
\line(10,0){447}
\vskip 1.1cm
\noindent
\noindent
{\large \bf Jan Troost}
\vskip 0.25cm
{\em 
\noindent
Laboratoire	 de	 Physique	 Th\'eorique \\
	 Département	 de	 Physique	 de	 l’ENS \\	 
École	 Normale	 Supérieure \\	 PSL	
Research	University,	Sorbonne	Universités,	UPMC,	CNRS \\ Paris, France
}
\vskip 1.2cm

\vskip0cm

\noindent
{\sc Abstract: } {We clarify three aspects of non-compact elliptic genera.
  Firstly, we give a path integral derivation of the elliptic genus of
  the cigar conformal field theory from its non-linear sigma-model
  description. The result is a manifestly modular sum
  over a lattice. Secondly, we discuss supersymmetric quantum
  mechanics with a continuous spectrum. We regulate the theory and
  analyze the dependence on the temperature of the trace weighted by the fermion number.
 The dependence is dictated by the
  regulator. From a detailed analysis of the dependence on the
  infrared boundary conditions, we argue that in non-compact elliptic genera right-moving
  supersymmetry combined with modular covariance is anomalous.
  Thirdly, we further clarify the relation between the flat space
  elliptic genus and the infinite level limit of the cigar elliptic
  genus.}

\vskip 1cm

\pagebreak

\newpage
\setcounter{tocdepth}{2}

\tableofcontents

\section{Introduction}
Mock modular forms have an illustrious history in mathematics \cite{Zagier}. However, a systematic understanding
of mock modular forms is recent \cite{Zwegers:2008zna} and evolving. Mock modular forms 
also appeared in physics
in various guises \cite{Eguchi:1987sm,Semikhatov:2003uc,Manschot:2009ia}. A 
natural habitat for mock modular forms and their non-holomorphic modular completion was provided by the
demonstration that they arise as elliptic genera of two-dimensional 
superconformal field theories with continuous spectrum \cite{Troost:2010ud}. As such the completed
forms appear also as duality covariant counterparts to black hole entropy counting functions
\cite{Dabholkar:2012nd}.

In this paper, we wish to clarify three aspects of non-compact
elliptic genera. The first comment we make is on the compact form of
the elliptic genus of the cigar derived by Eguchi and Sugawara in
\cite{Eguchi:2014vaa}.  It is a modular covariant sum over lattice
points which is an exponentially regulated Eisenstein series. Since it
is manifestly modular covariant, one can wonder whether it has a
simple direct path integral derivation. We demonstrate that a path
integration of the non-linear sigma-model description of the cigar
provides such a derivation.  The second remark, in section
\ref{SQMhalf}, is based on an analysis of the temperature dependence
of the weighted trace $Tr (-1)^F e^{- \beta H}$ in supersymmetric
quantum mechanics with a continuous spectrum. Upon regularization, the
trace becomes $\beta$-dependent in a manner that hinges upon the
choice of regulator.  We demonstrate this in detail, analyze the
supersymmetric regulator and its path integral incarnation, and the
role of infrared boundary conditions. We use it to lay bare the
unresolvable tension between right-moving supersymmetry and modularity
in the non-compact elliptic genus.  In a third and final part, we
clarify the relation between the flat space superconformal field
theory and the infinite level limit of the cigar conformal field
theory using their elliptic genera.

\section{The Path Integral Lattice Sum}
\label{lattice}
In this section, we wish to obtain a simpler path integral understanding of the compact formula
for the elliptic genus of the cigar in terms of a lattice sum, derived in \cite{Eguchi:2014vaa}. To that end, we
 provide a new derivation of the elliptic genus of the cigar, through its supersymmetric non-linear sigma-model description.
The latter has the advantage of being parameterized in terms of the physical degrees of freedom only.
\subsection{The Guises of the Genus}
The cigar elliptic genus 
\begin{equation}
\chi_{cig}(\tau,\alpha) 
= Tr_{RR} (-1)^{F_L+F_R} e^{2 \pi i  \alpha Q} q^{L_0-\frac{c}{24}} \bar{q}^{\bar{L}_0-\frac{c}{24}}
\label{egtrace}
\end{equation}
is a partition sum in the Ramond-Ramond sector, weighted by left- and right-moving fermion numbers $F_{L,R}$,
as well as twisted by the left-moving R-charge $Q$.
It was computed manifestly covariantly through a path integral over maps from the torus
into the coset $SL(2,\mathbb{R})/U(1)$ target space \cite{Troost:2010ud}. 
The result obtained in 
\cite{Troost:2010ud,Eguchi:2010cb,Ashok:2011cy} was
\begin{equation}
\chi_{cig}(\tau,\alpha) = k \int_0^1 ds_{1,2} \sum_{m,w \in \mathbb{Z}} \frac{\theta_{1}(s_1 \tau+s_2-\alpha \frac{k+1}{k},\tau)}{\theta_{1}(s_1 \tau+s_2-\frac{\alpha}{k},\tau)}
e^{2 \pi i \alpha w} e^{- \frac{k \pi}{\tau_2} |(m+s_2)+(w+s_1)\tau|^2} \, ,
\label{cigareg}
\end{equation}
where the $\theta_{1}$ functions arise from partition functions of
 fermions and bosons with twisted boundary conditions on the torus, the integers $m,w$ are winding numbers for the maps
from the torus onto the target space angular direction, and the angles $s_{1,2}$
are holonomies on the torus for the $U(1)$ gauge field used to gauge
an elliptic isometry of $ SL(2,\mathbb{R})$. The twist with respect to
the left-moving R-charge is given by $\alpha$.  This modular
Lagrangian result was  put into a Hamiltonian form in which the
elliptic genus could be read directly as a sum over right-moving
ground states plus an integral over the differences of spectral
densities for the continuous spectrum of bosonic and fermionic
right-movers \cite{Troost:2010ud,Ashok:2011cy} . The difference of
spectral densities is determined by the asymptotic supercharge
\cite{Akhoury:1984pt,Troost:2010ud,Ashok:2013kk}.

In \cite{Eguchi:2014vaa}, a rewriting of the  result (\ref{cigareg}) in terms of a lattice sum was obtained.
The resulting expression for the cigar elliptic genus is
\begin{equation}
\chi_{cig}(\tau,\alpha) = \frac{\theta_{1}(\alpha,\tau)}{2 \pi \eta^3}
\sum_{m,w \in \mathbb{Z}} \frac{e^{- \frac{\pi}{k \tau_2} (\alpha^2+|m-w \tau|^2+2 \alpha(m-w \bar{\tau}))}}{\alpha+m-w \tau}
\, . \label{latticesum}
\end{equation}
This expression is also  manifestly modular covariant, because it is written as
a sum over a lattice $\mathbb{Z}+\mathbb{Z} \tau$. Our goal in
this section is to understand the formula (\ref{latticesum}) in
a more direct manner than through the route laid out in
\cite{Eguchi:2014vaa,Troost:2010ud,Eguchi:2010cb,Ashok:2011cy}. We recall that a key
step in the derivation of the lattice sum (\ref{latticesum}) was to
first compute the elliptic genus of the infinite cover of the
$\mathbb{Z}_k$ orbifold of the trumpet geometry
\cite{Sugawara:2011vg,Eguchi:2014vaa}.
\subsection{The Infinite Cover of The Orbifolded Trumpet}
We start our calculation from  the cigar
geometry \cite{Elitzur:1991cb,Mandal:1991tz,Witten:1991yr}
\begin{eqnarray}
ds^2 &=& \alpha' k ( d \rho^2 + \tanh^2\rho \, d \theta^2)
\nonumber \\
e^{\Phi} &=& e^{\Phi_0}/\cosh \rho \, ,
% \nonumber \\
% \Delta &=& - \frac{1}{e^{-2 \Phi} \sqrt{G}}
% \partial_\mu (e^{- 2 \Phi} \sqrt{G} G^{\mu \nu} \partial_\nu)
% \nonumber \\
% G &=& (\alpha' k)^2 \tanh^2 \rho
% \nonumber \\
% \sqrt{G} &=& \alpha' k \tanh \rho
% \nonumber \\
% e^{-2 \Phi} &=& e^{- 2 \Phi_0} \cosh^2 \rho
% \nonumber \\
% \Delta &=& - \frac{1}{\alpha' k} ( \frac{1}{\sinh \rho \cosh \rho} \partial_\rho ( \sinh \rho \cosh \rho \partial_\rho)
% + \coth^2 \rho \partial_\theta^2)
% \nonumber \\
% &=& - \frac{1}{\alpha' k} ( \partial_\rho^2 + (\tanh \rho + \coth \rho) \partial_\rho
% + \coth^2 \rho \partial_\theta^2)
% \nonumber \\
% \Psi &=& e^{i n \theta} \sinh^{|n|} \rho F(..; - \sinh^2 \rho)
\end{eqnarray}
where the angle $\theta$ is identified modulo $2 \pi$. The metric and dilaton determine the couplings of a conformal two-dimensional
non-linear sigma-model.  The T-dual geometry is the $\mathbb{Z}_k$ orbifold of the trumpet:
\begin{eqnarray}
ds^2 &=& \alpha' (k d \rho^2 + \frac{1}{k} \coth^2 \rho \, d \theta^2)
\nonumber \\
e^{\Phi} &=& e^{\Phi_0}/\sinh \rho
% \Delta &=& - \frac{1}{e^{-2 \Phi} \sqrt{G}}
% \partial_\mu (e^{- 2 \Phi} \sqrt{G} G^{\mu \nu} \partial_\nu)
% \nonumber \\
% G &=& (\alpha' k)^2 \coth^2 \rho
% \nonumber \\
% \sqrt{G} &=& \alpha' k \coth \rho
% \nonumber \\
% e^{-2 \Phi} &=& e^{- 2 \Phi_0} \sinh^2 \rho 
% \nonumber \\
\end{eqnarray}
where the angle $\theta$ is again identified modulo $2 \pi$. The infinite cover
of the orbifold of the trumpet is the geometry in which we no longer impose any equivalence relation on the variable
$\theta$. 

We perform the path integral on the cover as follows. Firstly, we consider the
integral  over the zero modes and the oscillator modes separately. We suppose that the oscillator
contribution on the left is proportional to the free field result
\begin{equation}
Z_{osc}^{\infty} = \frac{1}{4 \pi^2 \tau_2} \frac{\theta_{1}(\alpha,\tau)}{ \eta^3} \, ,
\label{oscillatorcontribution}
\end{equation}
for a left-moving fermion of R-charge $1$ and two uncharged bosonic fields. The factor
$1/(4 \pi^2 \tau_2)$ is the result of the integral over momenta (at $\alpha'=1$). The right-moving oscillators cancel among
each other.

We want to focus on the remaining integral over zero modes, which contains
the crucial information on the modularly completed Appell-Lerch
sum \cite{Zwegers:2008zna}. The left-moving fermionic zero modes have been lifted by the R-charge twist. Thus, we can
concentrate on the integration over the bosonic zero modes as well as
the right-moving fermionic zero modes, with measure 
\begin{equation}
 d \rho d \theta d \tilde{\psi}^\rho d \tilde{\psi}^\theta \, .
\end{equation}
The square root of the determinant in the diffeomorphism invariant measures has canceled
between the bosons and the fermions. 
The relevant action is the $N=(1,1)$ supersymmetric extension of the non-linear sigma-model on the curved 
target space.\footnote{See e.g. formula (12.3.27) in \cite{Polchinski:1998rr}.} 
The  term in the action that lifts the right moving fermion zero modes is \cite{Polchinski:1998rr}
\begin{eqnarray}
S_{lift} &=& \frac{1}{4 \pi} \int d^2 z \, G_{\mu \nu} \tilde{\psi}^\mu \Gamma^\nu_{\rho \sigma} {\partial} X^\rho
\tilde{\psi}^\sigma
\label{liftterm}
\end{eqnarray}
and more specifically, the term proportional to the Christoffel connection symbols
\begin{eqnarray}
\Gamma_{\theta \theta \rho}=-\Gamma_{\rho \theta \theta} &=& \frac{1}{2}  \partial_\rho G_{\theta \theta} \, .
\label{Christoffel}
\end{eqnarray}
This leads to a term in the action equal to 
\begin{eqnarray}
S_{lift} &=& \frac{1}{4 \pi} \int d^2 z \, \tilde{\psi}^\theta \tilde{\psi}^\rho \partial_\rho G_{\theta \theta}  {\partial} \theta\, . \label{lift}
\end{eqnarray}
We can descend this term once from the exponential in order to absorb the right-moving zero modes and obtain 
a non-zero result.

We wish to introduce a twist in the worldsheet time direction for the target space angular direction
$\theta$ because we insert a R-charge twist operator in the elliptic genus, and the field $\theta$ is charged
under the R-symmetry \cite{Troost:2010ud,Eguchi:2010cb,Ashok:2011cy,Eguchi:2014vaa}. We thus must twist
\begin{equation}
\theta(\sigma_1+2 \pi \tau_1,\sigma_2+2 \pi \tau_2) = 
 \theta(\sigma_1,\sigma_2) +2 \pi \alpha \, ,
\end{equation}
and we still have $\theta(\sigma_1+ 2 \pi) = \theta(\sigma_1)$. Since we study the infinite cover of the $\mathbb{Z}_k$ orbifold
of the trumpet, there are no winding sectors.
We thus obtain the classical configuration
\begin{equation}
\theta_{cl} = \sigma_2 \alpha /\tau_2 \, . \label{thetaclassical1}
\end{equation}
We plug this classical solution (\ref{thetaclassical1}) into the action for the infinite order orbifold of the trumpet, and descend
a single insertion of (\ref{lift}) to lift the right-moving zero mode, use the Christoffel connection
(\ref{Christoffel}) and then find the zero mode integral
\begin{eqnarray}
Z^\infty_0 &=& 2 \pi N_\infty \int_0^\infty d \rho  \, \alpha \partial_\rho ( -\frac{\pi}{k}  \coth^2 \rho)
e^{- \frac{\pi \alpha^2}{ k \tau_2} \coth^2 \rho  }
\nonumber \\
&=& 2 \pi N_\infty  \frac{\tau_2}{\alpha}  e^{-  \frac{\pi \alpha^2}{ k \tau_2}} \, .
\label{zeromodefactor}
\end{eqnarray}
We have represented the integral over the variable $\theta$ by a factor of $2 \pi N_{\infty}$ where we think
of $N_{\infty}$ as the order of the cover, which goes to infinity.
Putting this together with the oscillator factor (\ref{oscillatorcontribution}) we proposed previously, we find
\begin{equation}
Z^\infty = N_{\infty} \frac{\theta_{1}(\alpha,\tau)}{\eta^3}
\frac{1}{2 \pi \alpha} e^{- \frac{\pi \alpha^2}{ k \tau_2} } \, .
\end{equation}
This precisely agrees with the elliptic genus of the infinite cover of
the orbifolded trumpet calculated in
\cite{Eguchi:2014vaa}.\footnote{The factor $N_{\infty}$ is absorbed in
  the definition of $Z^\infty$ in \cite{Sugawara:2011vg,Eguchi:2014vaa}.}
\subsection{The  Lattice Sum}
Our next step is the path integral incarnation of the procedure of the derivation of the lattice
sum formula in \cite{Eguchi:2014vaa}. We undo the infinite order orbifold of the cigar,
i.e. we undo the infinite order cover of the orbifolded trumpet. This will reproduce the lattice
sum elliptic genus formula.

There are two changes that we need to carefully track. The first one is that since the field
$\theta$ becomes an angular variable with period $ 2 \pi$, we must sum over the world sheet
winding sectors. Thus, we introduce the identifications
\begin{eqnarray}
\theta(\sigma_1+2 \pi \tau_1,\sigma_2+2 \pi \tau_2) &=& 
 \theta(\sigma_1,\sigma_2) + 2 \pi (\alpha+m) 
\nonumber \\
\theta(\sigma_1+ 2 \pi,\sigma_2) &=& \theta(\sigma_1,\sigma_2) + 2 \pi w \, ,
\end{eqnarray}
which lead to the classical solutions
\begin{eqnarray}
\theta_{cl} &=& \sigma^1 w + \sigma^2 (m+\alpha-w \tau_1)/\tau_2
\nonumber \\
%&=& \frac{z+ \bar{z}}{2} w + \frac{z-\bar{z}}{2i} (m+\alpha-w \tau_1)/\tau_2
%\nonumber \\
&=& \frac{-i}{2 \tau_2} (z (m+\alpha-w \bar{\tau})-\bar{z} (m+\alpha-w \tau)) \, .
\end{eqnarray}
We then have the classical contribution to the action
\begin{eqnarray}
\partial \theta_{cl} \bar{\partial} \theta_{cl} &=&
\frac{1}{4 \tau_2^2} (m+\alpha-w \bar{\tau}) (m+\alpha-w \tau)
\nonumber \\
&=& \frac{1}{4 \tau_2^2} (|\lambda|^2 + \alpha (\lambda+\bar{\lambda}+\alpha))
\label{exponent}
\end{eqnarray}
where $\lambda=m-w \tau$. After tracking normalization factors, one finds that the action acquires another overall factor of 
$4 \pi \tau_2/k$ (see e.g. \cite{Polchinski:1998rq}).

The second effect we must take into account is that the left-moving R-charge corresponds 
to the left-moving momentum of the angle field. When we introduce a winding number $w$, we must
properly take into account the contribution of the winding number to the left-moving momentum.
This amounts to adding a factor of $e^{-2 \pi i \alpha w/k}$ to a contribution arising from winding
number $w$. (Recall that the radius is $R^2/\alpha'=1/k$.)
We rewrite 
\begin{eqnarray}
e^{-2 \pi i \alpha w/k} % &=& e^{\alpha w (\tau-\bar{\tau}) \frac{\pi}{k \tau_2}}
%\nonumber \\
&=& e^{\alpha (\lambda-\bar{\lambda}) \frac{\pi}{k \tau_2} } \, 
\end{eqnarray}
which leads to a total contribution to the exponent equal to
\begin{equation}
- \frac{\pi}{k \tau_2} ( |\lambda|^2 + \alpha(\lambda+ \bar{\lambda}) +\alpha^2 + \alpha (-\lambda+\bar{\lambda}))
=- \frac{\pi}{k \tau_2}( |\lambda|^2 + 2 \alpha \bar{\lambda} + \alpha^2) \, .
\end{equation}
%This is the desired exponent.
The denominator in the final expression is obtained from a factor $(\lambda+\alpha)(\bar{\lambda}+\alpha)$ in the denominator
that arises from the exponent (\ref{exponent}) in the generalized zero mode integral (\ref{zeromodefactor}) on the one hand,  and
a factor of $\bar{\lambda}+\alpha$ in the numerator from the 
$z$-derivative of the angular variable $\theta$ on the other hand (arising from the zero mode lifting term (\ref{lift})). Multiplying these, we find the final formula
\begin{equation}\chi_{cig}(\tau,\alpha) = \frac{\theta_{1}(\alpha,\tau)}{2 \pi \eta^3}
\sum_{m,w \in \mathbb{Z}} \frac{e^{- \frac{\pi}{k \tau_2} (\alpha^2+|m-w \tau|^2+2 \alpha(m-w \bar{\tau}))}}{\alpha+m-w \tau}
\, , \label{latticesum2} 
\end{equation}
which is the compact lattice sum form \cite{Eguchi:2014vaa} of the cigar elliptic genus. We have
given a direct derivation of the lattice sum form, using the
non-linear sigma model description. This concludes the first
panel of our triptych.

\section{Supersymmetric Quantum Mechanics on a Half Line}
\label{SQMhalf}
In this section, we wish to render the fact that the non-holomorphic
term in non-compact elliptic genera arises from a contribution due to
the continuum of the right-moving supersymmetric quantum mechanics \cite{Troost:2010ud}
even more manifest.  For that purpose, we discuss to what extent
the right-moving supersymmetric quantum mechanics can be regularized
in a supersymmetric invariant way, or a modular covariant manner, but
not both. That fact leads to the holomorphic anomaly \cite{Troost:2010ud}.
The plan of this section is to first review  how boundary conditions in ordinary quantum mechanics
show up in its path integral formulation. We then extend this insight to supersymmetric quantum
mechanics. We illustrate the essence of the phenomenon in the simplest of systems.
We end with a discussion of how the  regulator of the non-compact elliptic genus cannot be both modular and supersymmetric, which leads to an anomaly.
\subsection{Quantum Mechanics on a Half Line}
\label{QM}
We are used to path integrals that map spaces with boundaries into closed manifolds. Less frequently,
we are confronted with path integrals from closed
spaces to spaces with boundaries. It is the latter case that we study in the following in the very
simple setting of quantum mechanics.

In particular, we discuss quantum mechanics on a half line, its path
integral formulation, and pay particular attention to the path
integral incarnation of the boundary conditions. The easiest way to
proceed will be to relate the problem to  quantum mechanics 
on the whole real line. What follows is a review of the results
derived in e.g. \cite{Clark:1980xt,Farhi:1989jz,Grosche:1993jr}, albeit from an original perspective.
\subsubsection{Quantum Mechanics on the Line}
Firstly, we rapidly  review  quantum mechanics on the real line. We work with a Hilbert space which consists
of quadratically integrable functions on the line parameterized by a coordinate $x$. We have a Hamiltonian operator $H$ of the form
\begin{equation}
H = - \frac{1}{2}  \partial_x^2 +  V(x) \, ,
\label{QMHamiltonian}
\end{equation} 
where $V(x)$ is a potential. We can define a Feynman amplitude to go from an initial position $x_i$ to a final position $x_f$ in time $t$ through the path
integral
\begin{equation}
A(x_i,x_f,t) = \int_{x(0)=x_i}^{x(t)=x_f} dx \, e^{iS[x]} \, ,
\end{equation}
where the action is equal to 
\begin{equation}
S = \int_0^t dt'( \frac{\dot{x}^2}{2}-V(x) ) \, .
\end{equation}
The Schr\"odinger equation for the wave-function of the particle reads
\begin{equation}
i \partial_t \Psi = H \Psi \, , \label{schrodinger}
\end{equation}
and we work with normalized wave-functions $\Psi$.
We can also write the amplitude in terms of an integral over energy eigenstates $\Psi_E$:
\begin{equation}
A(x_i,x_f,t) = \int dE e^{-iEt} \Psi_E(x_i) \Psi_E(x_f) \, , \label{Green}
\end{equation}
and the amplitude satisfies the $\delta$-function completeness relation at $t=0$, as well as the Schr\"odinger equation (\ref{schrodinger})
in the initial and final position variables $x_{i}$ and $x_{f}$.
\subsubsection{Quantum Mechanics on the Half Line}
The subtleties of quantum mechanics on the open real half line $x \ge 0$ have been understood
for a long time \cite{RS}. Boundary conditions compatible with unitarity have been classified.
 The path integral formulation for quantum mechanics
on the half line has resurfaced several times over the last decades \cite{Clark:1980xt,Farhi:1989jz,Grosche:1993jr},
and is also well-understood. We review what is known.

The half-line has a boundary, and we must have that the probability current vanishes at the boundary. This is
guaranteed by the Robin boundary conditions 
\begin{equation}
\partial_x \Psi(0) =  c \,  \Psi(0) \, . 
\end{equation}
When the constant $c$ is zero, we have a Neumann boundary condition and
when it is infinite, the boundary condition is  in effect Dirichlet, $\Psi(0)=
0$. Suppose we are given a Hamiltonian $H$ of the form
(\ref{QMHamiltonian}) with a potential $V(x)$ on the half line $x>0$. We can
extend the quantum mechanics on the half line to the whole real line
by extending the potential in an even fashion, declaring that
$V(-x)=V(x)$. It is important to note that this constraint leaves the
potential to take any value at the origin $x=0$. We can then think of
the quantum mechanics on the half line as a folded version of the
quantum mechanics on the real line.\footnote{In string theory, one
  would say that we think of the half line as an orbifold of the real
  line.}  The even quantum mechanics that we constructed  on the real line has a global symmetry
group $\mathbb{Z}_2$. We can divide the quantum mechanics problem on the real line,
including its Hilbert space, by the $\mathbb{Z}_2$ operation, and find
a well-defined quantum mechanics problem on the half line, which is
the original problem we wished to discuss.

An advantage of this way of thinking is that the measure for quantum
mechanics on the whole line is canonical. It leads to the Green's
function (\ref{Green}). Since the quantum mechanics
that we constructed has a global $\mathbb{Z}_2$ symmetry, we can 
classify eigenfunctions in terms of the representation they form under
the $\mathbb{Z}_2$ symmetry, namely, we can classify them into even
and odd eigenfunctions of the Hamiltonian. We then obtain the whole
line Green's function in the form that separates the even and odd
energy eigenfunction contributions
\begin{equation}
A(x_i,x_f,t) = \int dE e^{-iEt} (\Psi_{E,e}(x_i) \Psi_{E,e}(x_f)+\Psi_{E,o}(x_i) \Psi_{E,o}(x_f) )\, .
\end{equation} 
The Green's function
\begin{equation}
A^{\frac{1}{2},D}(x_i,x_f,t) = \frac{1}{2}(A(x_i,x_f,t) -A(x_i,-x_f,t)) =  \int dE e^{-iEt} \Psi_{E,o}(x_i) \Psi_{E,o}(x_f) \, ,
\end{equation}
is well-defined on the half-line and satisfies Dirichlet boundary
conditions. We divide by a factor of two since we are projecting onto
$\mathbb{Z}_2$ invariant states. {From} the path integral perspective,
the subtraction corresponds to a difference over paths that go from
$x_i$ to $x_f$ and that go from $x_i$ to $-x_f$, on the whole real
line, with the canonical measure (divided by two). This prescription generates a
measure on the half line which avoids the origin, since we subtract
all paths that cross to the other side
\cite{Clark:1980xt,Farhi:1989jz}.\footnote{This is a common
  manipulation in probability theory.} If we represent the
$\mathbb{Z}_2$ action oppositely on the odd wave-functions, we arrive
at the Green's function that satisfies Neumann boundary conditions:
\begin{equation}
A^{\frac{1}{2},N}(x_i,x_f,t) = \frac{1}{2}( A(x_i,x_f,t) +A(x_i,-x_f,t)) =  \int dE e^{-iEt} \Psi_{E,e}(x_i) \Psi_{E,e}(x_f) \, .
\end{equation}
In this second option, we add paths to the final positions
$x_f$ and $-x_f$ with their whole line weights (divided by
two). This path integral represents a sum over paths that reflect an
even or an odd number of times off the origin $x=0$, and in
particular, allows the particle to reach the end of the half line.

We clearly see that the naive folding operation projects the states of
the quantum mechanics onto those states that are even, or those that
are odd.\footnote{These are states in the untwisted sector of an
  orbifold, projected onto invariants under the gauged discrete
  symmetry.} However, concentrating on these two possibilities only fails
to fully exploit the loop hole that the even potential $V(x)$ allows, which
is an arbitrary value $V(0)$ at the fixed point $x=0$ of the folding
operation.\footnote{In string theory orbifolds, the fixed point hosts
  extra degrees of freedom which in that case are very strongly
  constrained by consistency.}
We can make use of this freedom by taking as the total potential an even potential
$V(x)$, zero at $x=0$, complemented with a $\delta$-function:
\begin{equation}
H^c(x) = -\partial_x^2/2 + V(x) + c \, \delta(x) \, .
\end{equation}
We take the wave-function on the whole line to be even and continuous, with a discontinuous
first derivative at the origin. When we consider the one-sided derivative at zero, we find that the 
wave-function satisfies the Robin boundary condition \cite{Farhi:1989jz}
\begin{equation}
\partial_x \Psi(0^+) = c \Psi(0) \, .
\end{equation}
We have gone from a purely even continuous and differentiable
wave-function on the real line  that satisfies the Neumann boundary
condition (at $c=0$) to an even wave-function that satisfies mixed Robin boundary
conditions, by influencing the wave-function near zero with a
delta-function interaction.\footnote{The even wave-function on the
  side $x>0$ corresponds to the linear combination $\Psi(x) \propto
  (\Phi_{E,e}(x)+c \, \Phi_{E,o}(x))$ in terms of even and odd solutions
  to the problem on the real line without the delta-function
  interaction \cite{Farhi:1989jz}. It is an invariant under the $\mathbb{Z}_2$ action with discontinuous
derivative at the origin.} It is intuitively clear, and argued in detail in
\cite{Farhi:1989jz} that it is harder to push an initial problem with
Dirichlet boundary conditions at the origin towards a mixed boundary
condition problem.  In order to achieve this, one needs a very deep
well \cite{Farhi:1989jz}. For later purposes, we note in particular that an ordinary
delta-function insertion at the origin will not influence an initial
Dirichlet boundary value problem.

As an intuitive picture, we can imagine that the delta-function is generated by possible extra
degrees of freedom that are localized at the origin, and whose interaction with the quantum mechanical degree of freedom
we concentrate on induces the delta-function potential localized at the origin.

Thus far, we briefly reviewed the results of
\cite{Clark:1980xt,Farhi:1989jz} on path integrals on the half line and discussed how they are
consistent with folding. Next, we render these  techniques
compatible with supersymmetry.

\subsection{Supersymmetric Quantum Mechanics on the Half Line}
\label{SQM}
In this section, we extend our perspective on quantum mechanics on the half line
to a quantum mechanical model with supersymmetry. We again start from a quantum mechanics on the whole of the
real line, with extra fermionic degrees of freedom and supersymmetry. In a second stage, we  fold
the quantum mechanics onto the half line in a manner consistent with supersymmetry.
\subsubsection{Supersymmetric Quantum Mechanics on the Line}
We discuss the supersymmetric system with
Euclidean action (see e.g. \cite{Cooper:1994eh})
\begin{eqnarray}
S_E &=& \int_0^t d \tau( \frac{1}{2} \partial_\tau x^2 + \frac{1}{2} W^2 - \psi^\ast (\partial_\tau - W')\psi) \, ,
\end{eqnarray}
where $W'(x)=\partial_x W(x)$.
%and the path integral
%\begin{equation}
%Z = \int d x d \psi d  \psi^\ast e^{-S_E}
%\end{equation}
The action permits two supersymmetries with infinitesimal variations
\begin{eqnarray}
\delta x &=& \epsilon^\ast \psi + \psi^\ast \epsilon
\nonumber \\
\delta \psi^\ast &=& - \epsilon^\ast (\partial_\tau{x}+W)
\nonumber \\
\delta \psi &=& \epsilon (  \partial_\tau x - W) \, . \label{susyvars}
\end{eqnarray}
When we quantize the fermionic degrees of freedom, we tensor the space of quadratically integrable functions with a two 
component system.  We call one component bosonic and the other
fermionic. The two components have the Hamiltonians \cite{Cooper:1994eh}\footnote{We follow
  standard conventions for supersymmetric quantum mechanics in this
  section. These differ by a factor of two from the standard conventions
  for quantum mechanics used in section \ref{QM}.}
\begin{equation}
H_\pm = p^2 + W^2  \mp  W' \, . \label{susyHs}
\end{equation}
We  introduced the operator 
\begin{equation}
p = -i \partial_x
\end{equation}
and can represent the supercharges by 
\begin{eqnarray}
Q &=& (p + i W) \left( \begin{array}{cc} 0 & 0 \\ 1 & 0 \end{array} \right)
\nonumber \\
Q^\dagger &=& (p-iW) \left( \begin{array}{cc} 0 & 1 \\ 0 & 0 \end{array} \right) \, .
\end{eqnarray}
 When we  trace over the fermionic degrees of freedom, we need to compute 
the fermionic determinant with  anti-periodic boundary conditions. It evaluates
to \cite{Cooper:1994eh}
\begin{equation}
Z_f^{anti-per}(x) = \int d \psi d \psi^\ast_{anti-per} \exp( \psi^\ast (\partial_\tau - W')\psi )
= \cosh (\int_0^t d \tau \frac{W'(x)}{2}) \, , \label{cosh}
\end{equation}
after regularization. This is the path integral counterpart to the calculation of the Hamiltonians
(\ref{susyHs}).

\subsubsection{Supersymmetric Quantum Mechanics on the Half Line}
We study the supersymmetric quantum mechanics on the half line
by folding the supersymmetric quantum mechanics on the whole line. We wish for the folding
$\mathbb{Z}_2$
symmetry to preserve supersymmetry. Since the particle position $x$ is
odd under the $\mathbb{Z}_2$ action (as is its derivative with respect
to time, since we choose world line time to be invariant), we demand that the
superpotential $W(x)$ is odd under parity, and that the fermionic
variables $\psi$ and $\psi^\ast$ are odd as well. See equation (\ref{susyvars}). Thus, we have the $\mathbb{Z}_2$ action
\begin{equation}
(x,\psi,\psi^\ast) \rightarrow (-x,-\psi,-\psi^\ast) \, ,
\label{susyZ2}
\end{equation}
and the superpotential $W$ is odd. For the moment, we consider the superpotential to be 
continuous, and therefore zero at zero.

We project onto states invariant under the $\mathbb{Z}_2$ action (\ref{susyZ2}).
Thus, in any path integral, we will insert a projection operator $P_{\mathbb{Z}_2}$ that consists
of 
\begin{equation}
P_{\mathbb{Z}_2} = \frac{1}{2} (1+P (-1)^F)
\label{projop}
\end{equation}
where $P$ is the parity operator that maps $P : x \rightarrow -x$ and $(-1)^F$ maps fermions to minus
themselves. 
When we trace over the fermionic degrees of freedom with a $(-1)^F$ insertion,
we must impose periodic boundary conditions on the fermions. The fermionic determinant in this
case evaluates to \cite{Cooper:1994eh}
\begin{equation}
Z_f^{per}(x)  = \int d \psi d \psi^\ast_{per} \exp( \psi^\ast (\partial_\tau - W')\psi )
= \sinh (\int_0^T d \tau \frac{W'(x)}{2}) \, ,
\end{equation}
which leads to the same Hamiltonians (\ref{susyHs}) for the two component system,
and when we compare to equation (\ref{cosh}) we find  a minus sign up front in the path integral
over the second component. 
As a consequence, for the first component of the two component system, from the insertion of the projection operator $P_{\mathbb{Z}_2}$ in equation (\ref{projop}), we will obtain
a path integral measure
\begin{eqnarray}
\frac{1}{2} (\int_{x_i}^{x_f} dx + \int_{x_i}^{-x_f} dx) \, ,
\end{eqnarray}
while for the second component, we obtain a path integral measure
\begin{eqnarray}
\frac{1}{2} (\int_{x_i}^{x_f} dx - \int_{x_i}^{-x_f} dx) \, .
\end{eqnarray}
Thus, from the discussion in subsection \ref{QM}, the upper component, which we will call fermionic and indicate
with a minus sign, will satisfy a Neumannn boundary condition at zero,
while the bosonic component will satisfy the Dirichlet boundary
condition. We carefully crafted our set-up to be consistent with
supersymmetry, and must therefore expect the boundary conditions we
obtain to be consistent with supersymmetry as well.  Indeed, the operator $Q$ maps the derivative of the fermionic
wave-function to the bosonic wave-function (when evaluated at the boundary, and using $W(0)=0$).
 Thus, the operator $Q$ maps the boundary conditions into one another.\footnote{
Note that the choice of action of $(-1)^F$ on the two components (assigning to one component a plus sign)
broke the symmetry between $Q$ and $Q^\dagger$ in this discussion. In other words, the opposite assignment would have
resulted in the operator $Q^\dagger$ mapping one boundary condition into the other.}
%XXX What about the supercharge $Q^\dagger$ ? XXX

The next case we wish to study is when the superpotential is well-defined on the half-line for 
$x>0$, and approximates a non-zero constant as we tend towards $x=0$. Since the superpotential is
odd on the line, the distributional derivative of the superpotential will be a delta-function
with coefficient twice the limit of the superpotential as it tends towards zero. If we call the latter
value $W_0$, then we have the equation 
\begin{equation}
W'(0)=2 W_0 \, \delta(x) \, .
\end{equation}
The derivative of the superpotential arises as a term in the component Hamiltonians
(\ref{susyHs}). The $\delta$-function interaction at the origin will result in a change in the Neumann (but not the Dirichlet) boundary conditions,
as we saw in subsection \ref{QM}. If we follow through the consequences, we find that the supersymmetric quantum
mechanics on the half line that we obtain by folding now satisfies the boundary conditions
\begin{eqnarray}
\Psi_+ (0) &=& 0
\nonumber \\
\partial_x \Psi_-(0) &=& W_0 \Psi_-(0) \, .
\end{eqnarray}
These boundary conditions are consistent with supersymmetry.

\subsubsection{An Interval}
We have used the folding technique to obtain a supersymmetric or
ordinary quantum mechanics problem on a half line.  We can use the
same technique to generate quantum mechanics problems on an
interval. We  perform a second folding by the reflection symmetry $x
\rightarrow 2L-x$ where $L$ is the length of the desired
interval. The fermions also transform with a minus sign under the second
$\mathbb{Z}_2$ generator. Again, we can render the superpotential odd under the second
flip, take into account a possible delta-function potential on the
second end of the interval, and find boundary conditions consistent with supersymmetry on both
ends. 
Our  application of these ideas lies in regulating a weighted trace,
and we  proceed immediately to apply them in that particular context.

\subsection{Infrared Regulators and the Weighted Trace}
We wish to discuss the trace 
\begin{equation}
Z(\beta) = Tr (-1)^F e^{- \beta H}
\end{equation}
over the Hilbert space of states, weighted with a sign $(-1)^F$ corresponding to their 
fermion number $F$. It is well-known that this
weighted trace is equal to the supersymmetric (Witten) index when the
spectrum of the supersymmetric quantum mechanics is discrete \cite{Witten:1982df}. It then
reduces to the index which equals the number of bosonic minus the
number of fermionic ground states.\footnote{We use the name weighted trace because we will
soon encounter contexts in which it is not an index.}

When the spectrum of the supersymmetric quantum mechanics is
continuous, the situation is considerably more complicated (see e.g. \cite{Niemi:1983ds,Akhoury:1984pt,Alves:1986tp}), 
and the debate in the literature  on this quantity may not have culminated
in a clear pedagogical summary.  We attempt to improve the state of affairs in this subsection.
The origin of the difficulties is that the trace over a continuum of states is an
ill-defined concept. An infinite set of states contributing a finite amount gives rise to a divergent sum.
 A proper definition requires a regulator. An infrared regulator will reduce the continuum to a discretuum and 
render the  trace finite. The alternating sum can remain finite in the limit where we remove the regulator. There has been a discussion on whether
and how the resulting weighted trace $Z(\beta)$ depends on the inverse temperature $\beta$, and on the infrared regulator.
To understand the main issues at stake, and to draw firm conclusions,  it is sufficient to consider
the  example of a free supersymmetric particle on the half line.

\subsubsection{The Free Supersymmetric Particle on the Half Line}

Let us consider a supersymmetric quantum mechanics, based on the superpotential
which is equal to a constant for $x>0$, namely $W(x>0)=W_0$. We obtain the half line supersymmetric
quantum mechanics by folding the problem on the whole line, and induce supersymmetric boundary conditions
at the end of the half line. We recall the Hamiltonians
\begin{equation}
H_\pm = p^2 + W_0^2 \mp 2  \, W_0 \, \delta(x) \, ,
\end{equation}
with  boundary conditions
\begin{eqnarray}
\partial_x \Psi_{-} &=& W_0 \Psi_{-}
\nonumber \\
\Psi_{+}(0) &=& 0 \, .
\end{eqnarray}
We can then solve for the wave-functions on the half line. The solutions for energy $E=p^2+W_0^2$ are given by
reflecting waves. The phase shift is set by the boundary condition. We have the wave-functions on the half line $x \ge 0$
\begin{eqnarray}
\Psi_+(x) &=& c_+ (e^{ipx}-e^{-ipx}) \, , 
\nonumber \\
\Psi_-(x) &=& c_- (e^{ipx} + \frac{ip - W_0}{ip+W_0} e^{-ipx}) \, .
\label{solns}
\end{eqnarray}
We find that the supercharge $Q$ maps the wave-function $\Psi_-$ into $\Psi_+$ if we identify $c_- (p+ i
W_0) = c_+$. Thus, we have computed the space of eigenfunctions for
bosons and fermions and how they are related.

\subsubsection{The Weighted Trace}
Our intermediate goal is to evaluate the weighted trace $Z(\beta)$ in
this model. To evaluate the trace, we need an infrared regulator.
Moreover, the weighted trace  depends on the infrared regulator, as
we will demonstrate. In any case, we need to  introduce an infrared
regulator to make the trace well-defined. We cut off the space at large $x=x_{IR}$. We need to impose
boundary conditions at this second end, at $x_{IR}$. As a result, the spectrum becomes
discrete, and we will be able to perform the trace over states
weighted by the corresponding fermion number. We consider two regulators in detail.

In a first regularization, we construct the supersymmetric quantum mechanics on the interval as we described previously.
The result will be a Hamiltonian 
\begin{equation}
H_\mp = p^2 + W_0^2 \pm  2  W_0 \,  \delta(x) \mp 2 W_0 \, \delta(x-x_{IR}) \, ,
\label{intervalH}
\end{equation}
and boundary conditions
\begin{eqnarray}
\partial_x \Psi_{f}(0^+) &=& W_0 \Psi_{f}
\nonumber \\
\Psi_{b}(0) &=& 0 
\nonumber \\
\partial_x \Psi_{f}(x_{IR}^-) &=& W_0 \Psi_{f} \nonumber \\
\Psi_b(x_{IR}) &=& 0\, .
\end{eqnarray}
The reason that the boundary condition on both sides is the same
despite the sign flip in the $\delta$ function coefficient in (\ref{intervalH}) is because
we are evaluating either the derivative with a left or a right
approach to the singular point. 
Because the $\mathbb{Z}_2 \times \mathbb{Z}_2$ folding procedures commute
with supersymmetry,  the infrared
regulated model preserves supersymmetry. Explicitly, we have a spectrum
determined by the infrared boundary condition
\begin{equation}
e^{ip_n x_{IR}} -e^{-i p_n x_{IR}} = 0 \, ,
\end{equation}
which implies
\begin{equation}
p_n = \frac{\pi n}{x_{IR}}
\end{equation}
where $n$ is an integer. All states are two-fold degenerate. The state with the lowest energy has energy equal
to $E=W_0^2$. The weighted trace reduces to a supersymmetric index and the Witten index is equal to zero.

A second regularization of the weighted trace proceeds as follows. We
rather put  Dirichlet boundary conditions at the infrared cut-off
$x_{IR}$ for both component wave-functions. We can intuitively argue that we expect a normalizable
wave-function to drop off at infinity, and that the Dirichlet boundary
condition is a good approximation to this expectation. It has the
added advantage of not introducing extra degrees of freedom at the end
point which we imagine to be responsible for a delta-function
potential. The disadvantage is that this infrared regulator breaks supersymmetry.
The regulated weighted trace will now sum over bosonic and fermionic states determined by the respective
conditions (see (\ref{solns}))
\begin{eqnarray}
e^{ip_n^b x_{IR}} -e^{i p_n^b x_{IR}} &=& 0 \, ,
\nonumber \\
e^{i p_{n'}^f x_{IR}} + \frac{ip_{n'}^f-W_0}{ip_{n'}^f+W_0} e^{-i p^f_{n'} x_{IR}} &=& 0 \, .
\end{eqnarray}
We define the phase shift
\begin{equation}
e^{i \delta(p)} = \frac{ip + W_0}{ip - W_0}
\end{equation}
of the fermionic wave-function. Then the solutions to the bosonic and fermionic boundary conditions are
\begin{eqnarray}
p_n^b &=& \frac{\pi n}{x_{IR}}
\nonumber \\
2 p^f_{n'} x_{IR} + \delta(p^f_{n'}) &=& 2 \pi (n' + \frac{1}{2}) \, .
\end{eqnarray}
As the infrared cut-off is taken larger, the number of states per
small $dp$ interval will grow, to finally reach the continuum we
started out with. To measure this growth, we can compute the bosonic and fermionic densities of states
\begin{eqnarray}
\rho^b(p) &=& \frac{dn}{dp} = \frac{x_{IR}}{\pi}
\nonumber \\
\rho^f(p) &=& \frac{dn'}{dp} = \frac{1}{2 \pi} (2 x_{IR}+ \frac{d \delta(p)}{dp}) \, .
\end{eqnarray}
Thus, when we approximate the weighted trace at large infrared cut-off by the appropriate integral formula,
we find \cite{Akhoury:1984pt}
\begin{equation}
Tr (-1)^F e^{- \beta H} = \int_0^\infty dp (\rho^b(p)-\rho^f(p)) e^{- \beta E(p)}
\end{equation}
where the difference of densities of states is given by
\begin{eqnarray}
\Delta \rho &=& \rho^b(p)-\rho^f(p)
= \frac{1}{2 \pi} \delta'(p) \nonumber \\
&=&  \frac{1}{2 \pi i} \frac{d}{dp} \log  \frac{ip + W_0}{ip - W_0}
= \frac{1}{2 \pi } (\frac{1}{ip + W_0} - \frac{1}{ip -W_0}) \, .
\end{eqnarray}

This second way of regularizing shows that the boundary condition we
impose at the infrared end of our interval is crucial in determining the
end result. When we put, as we did in the first case, a boundary
condition consistent with supersymmetry, then the difference of
spectral densities is zero for all values of the cut-off, and therefore also in the limit
of infinite cut-off. When we put identical boundary conditions
for  fermions and bosons at the infrared endpoint, then the
spectral densities differ by the phase shift in the continuum
problem. It should now be clear that one can choose another mix of
boundary conditions that will lead to yet another outcome for the
spectral measure. Before a choice of regulator, the weighted trace is
ill-defined. The final result depends on the regulator choice, even
after we remove the regulator. We have illustrated this effect in two
cases, but there is an infinite number of choices, and the
$\beta$-dependence of the final result $Z(\beta)$ is determined by
the choice of regulator. We should rather think of the weighted trace $Z(\beta,\mbox{regulator})$
as a function of both the inverse temperature $\beta$ and the regulator.

The first regulator is interesting, since it preserves
supersymmetry. The second regulator, with identical boundary
conditions for bosons and fermions is also interesting, it turns
out. Although we computed the spectral density in our particular model
of the free particle on a half line, the final result is
universal in an appropriate sense. The relative phase shift of bosons and fermions at large
$x_{IR}$ is determined by the asymptotic form of the supercharge $Q$
alone. This can be seen from the fact that the fermionic wave function in the infrared is determined by the bosonic wave function
in the infrared and the asymptotic supercharge.
 Thus, only the asymptotic value of the superpotential $\lim_{x
  \rightarrow \infty} W(x)=W_0$, which we assume to be constant, will
enter the phase shift and spectral density formula \cite{Akhoury:1984pt}. Thus, the result
for the $\beta$-dependent weighted trace is universal, {\em given} the regularization procedure.
Both the universality and the caveat are crucial.

The final result for our free particle on the half line with Dirichlet infrared regulator becomes  \cite{Akhoury:1984pt}
\begin{eqnarray}
Z(\beta,\mbox{Dirichlet}) &=& \int_0^\infty dp  \frac{1}{2 \pi } (\frac{1}{ip + W_0} - \frac{1}{ip -W_0}) 
e^{-\beta (p^2 + W_0^2)}
\nonumber \\
&=&  \int_{- \infty}^{+\infty} dp  \frac{1}{2 \pi } \frac{1}{ip + W_0} 
e^{-\beta (p^2 + W_0^2)}
%\nonumber \\
%&=&\frac{1}{2}(-1+\mbox{sgn}(W_0)+ \mbox{erfc}(\sqrt{\beta} W_0)) 
\, .
\end{eqnarray}
%XXX Some discrepancy still with Akhoury-Comtet ? XXX

\subsubsection*{Conclusion}
Of course, we recuperated the standard wisdom that any supersymmetric regulator makes the weighted trace into a
supersymmetric Witten index which is $\beta$-independent. However, another choice of infrared
regulator can give rise to a $\beta$-dependent weighted trace, and the $\beta$-dependence 
is dictated by the regulator.

It is quite striking that there are applications of
supersymmetric quantum mechanics on a half line in which the infrared
regulator is dictated by another symmetry of an overarching, higher
dimensional model. In such circumstances, the weighted trace and its
$\beta$-dependence become  well-defined and  useful
concepts.

\subsection{The Application to the Elliptic Genus}
In the calculation of the cigar elliptic genus (\ref{egtrace}), there is a weighted
trace over the right-moving supersymmetric quantum mechanics. For each
sector labeled by the right-moving momentum $\bar{m}$ on the
asymptotic circle of the cigar, there is a supersymmetric quantum mechanics with
superpotential $W$ that asymptotes to $W_0=\bar{m}$ \cite{Ashok:2013kk}. The point is now
that, as we saw, each of the right-moving supersymmetric quantum
mechanics labeled by the right-moving momentum can be cut-off
supersymmetrically using a $\delta$-function potential with
coefficient depending on the right-moving momentum $\bar{m}$. The
resulting elliptic genus would be equal to the mock modular Appell-Lerch sum.
 The cut-off depending on the right-moving momentum is not
modular covariant though.  The right-moving momentum is a combination
of a winding number of torus maps, and the Poisson dual of the other
winding number of torus maps, and as a result does not transform modular covariantly.  The
second alternative (and the one generically preferred in the context
of a two-dimensional theory of gravity in which we wish to preserve
large diffeomorphisms as a symmetry group) is to have a Dirichlet cut-off for
all these supersymmetric quantum mechanics labeled by the
right-moving momentum. This choice {\em is} covariant under modular
transformations, but is not supersymmetric, as we have shown. The
result of the second regularization is a modular completion of the mock modular form. 
We have thus shown that an anomaly arises in the combination of right-moving supersymmetry and
modular covariance.

Our analysis of supersymmetric quantum mechanics is interesting in
itself.  It also provides the technical details of the reasoning in
\cite{Troost:2010ud,Ashok:2011cy}, and thus produces a second panel in
our elliptic triptych.  Moreover, our technical tinkering paints the
background to continuum contributions to indices, or rather  their continuous
counterparts in two-dimensional theories \cite{Ashok:2014nua} as well
as in four-dimensional theories with eight supercharges
\cite{Alexandrov:2014wca,Pioline:2015wza}. In particular, it
clarifies both the regulator dependence as well as the universality of
the results on weighted traces in the presence of supersymmetry and a continuum.

\section{A Flat Space Limit Conformal Field Theory}
In \cite{Giveon:2014hfa}, we studied the infinite level limit of the cigar elliptic genus. In this limit, the target
space is flattened. One is tempted to interpret the resulting conformal field theory as a flat space
supersymmetric conformal field theory at central charge $c=3$. Still, the theory 
has features that distinguish it from a 
mundane flat space theory. 
In this third panel, we add remarks to the discussion provided in \cite{Giveon:2014hfa}, to which we also refer for
further context.

\subsection{Flat Space Regulated}
Firstly, we consider a flat space conformal field theory on $\mathbb{R}^2$, with two free bosonic scalar
fields, and two free Majorana fermions, for a total central charge of $c=3$, and with $N=(2,2)$
supersymmetry. We consider the Ramond-Ramond sector of the left- and right-moving  fermions.

%\subsubsection*{The Regulated Index}
The ordinary bosonic partition function is divergent. There is an overall volume factor arising from the integral
over bosonic zero modes which makes the partition function ill-defined. We can regulate the divergence in various
ways. One regulator would be to compactify the target space on a torus of volume $V$,
and then take the radii of the torus to infinity. The result is that the partition function approximates
(see e.g. \cite{Polchinski:1998rq})
\begin{equation}
Z_{V} = \frac{V}{\alpha'} (4 \pi^2  \tau_2)^{-1} | \eta|^{-4} \, ,
\label{volumeregulator}
\end{equation}
where $V/{\alpha'}$ represents the volume divergence. Alternatively,
we can compute the partition function through zeta-function regularization and the first Kronecker limit formula.
See e.g. \cite{DiFrancesco:1997nk}. The result is identical. If we regulate the bosons in this manner, and leave
the finite fermionic partition function unaltered, both the right-moving fermions and the left-moving fermions will
provide a zero mode in the Ramond-Ramond sector partition sum. Thus, we will find that the regulated supersymmetric Witten
index is zero for all finite values of the volume regulator $V$. The limit of the supersymmetric index will be zero
under these circumstances.

A  different way of regularizing is to twist the phase of the complex boson $Z=X^1+iX^2$. In the path integral
calculation of the complex boson partition function, this is implemented in a modular covariant way by 
demanding that the field configurations we integrate over pick up a phase as we go around a cycle of the torus.
The phase is a character of the $\mathbb{Z}^2$ homotopy group of the torus. If we parameterize the phases
by $e^{2 \pi i u m+ 2\pi i v w}$ (for winding numbers $m,w$ on the two cycles of the torus), the result can be obtained
either as the Ray-Singer analytic torsion \cite{Ray:1973sb} (to the power minus two) or by using the second Kronecker
limit formula. The modular invariant result is
\begin{equation}
Z_{twist} = |e^{-\frac{\pi (\textrm{Im}(\beta))^2}{\tau_2}} \frac{\theta_{1}(\beta,\tau)}{\eta}|^{-2} \, ,
\end{equation} 
where $\beta=u-v \tau$ is the complexified twist.
Near zero twist, there is a second order divergence that is proportional to $|\beta|^{-2} |\eta|^{-4}$ in accord with
equation (\ref{volumeregulator}).
The twist regulator breaks the translation invariance in space-time and preserves the rotational invariance. In fact,
it uses the rotation invariance to twist the angular direction and to remove all bosonic zero modes. (The idea is generic
in that one can use twists by global symmetries to lift divergences in numerous contexts.) If we leave the 
fermions undisturbed, we again have the fermionic zero modes that give rise to a zero elliptic genus for the full
conformal field theory.

The twist regulator suggests an interesting alternative. We
can twist the bosons and preserve world sheet supersymmetry at the
same time. The (tangent indexed) fermions naturally transform under the $SO(2)$ rotating
the two space-time directions, and if we twist with respect to the
complete action of the space-time rotations, we twist the fermions as
well. In that case, we find a partition function that equals one
\begin{equation}
Z_{twist} =|e^{-\frac{\pi (\textrm{Im}(\beta))^2}{\tau_2}} \frac{\theta_{1}(\beta,\tau)}{\eta}|^{2} \times |e^{-\frac{\pi (\textrm{Im}(\beta))^2}{\tau_2}} \frac{\theta_{1}(\beta,\tau)}{\eta}|^{-2}=1 \, .
\end{equation} 
 The
two fermionic zero modes have canceled the quadratic volume
divergence. The supersymmetric partition function (or Witten index) is
now equal to one for all values of the twist, and therefore equals one in the
limit where we remove the twist.

Again, as in section \ref{SQMhalf}, we see that the final result is regulator dependent (as is infinity times zero).
We have two regulators that preserve world sheet supersymmetry as well modular invariance, and they give rise
to index equal to zero, or to one.

\subsection{Twist Two}
We analyze how the above remarks influence our reading of the infinite
level limit of the cigar elliptic genus \cite{Giveon:2014hfa}. First off, we further twist the left-moving
fermions (only) by their left-moving R-charge, and  wind up with the
modular invariant flat space partition sum
\begin{eqnarray}
Z_{twist two} &=&  |\frac{ e^{-\frac{\pi (\textrm{Im}(\alpha+\beta))^2}{\tau_2}} \theta_{1}(\alpha+\beta,\tau)}{
 e^{-\frac{\pi (\textrm{Im}(\beta))^2}{\tau_2}}
\theta_{1}(\beta,\tau)}| \, . \label{twisttwo}
\end{eqnarray}
This chiral partition function suffers from a chiral anomaly. We have
again decided (for now) on a modular invariant choice of phase. The
regulating twist $\beta$ has canceled the right-moving zero mode
against the anti-holomorphic pole due to the infinite volume. The
left-moving R-charge twist $\alpha$ (when non-equivalent to zero) has 
reintroduced the holomorphic pole in $\beta$, also associated to the
divergent volume. When we take the limit $\beta \rightarrow 0$, we 
therefore again find an infinite result.

Once more, there are various ways to regularize the expression. One straightforward way to obtain the result
in \cite{Giveon:2014hfa} is to perform a modular covariant minimal subtraction. We expand the expression (\ref{twisttwo}) near $\beta=0$,
and subtract the pole. Given the dictum of a modular covariant transformation rule for the constant
term (e.g. the desired modular covariant transformation rule for the elliptic genus \cite{Kawai:1993jk}) one then obtains the
result \cite{Giveon:2014hfa}
\begin{equation}
Z_{ms,cov} = -\frac{1}{2 \pi} \frac{\partial_\alpha \theta_{11}(\alpha,\tau)}{\eta^3} - \frac{\alpha}{2 \tau_2}
\frac{\theta_{11}(\alpha,\tau)}{\eta^3} \, . \label{flatfinal}
\end{equation}
The cigar elliptic genus manages to regulate the pole at $\beta=0$ in
a more subtle manner than the covariant minimal subtraction advocated above \cite{Troost:2010ud}. It goes as follows.
One introduces an extra circle. Then, one
couples the circle to the angular direction of the plane (or the
cigar), and gauges a $U(1)$ such as to identify the two circular
directions. The net effect on the toroidal partition function is to
incorporate the twist $\beta$ into a modular covariant
holonomy  integral. 
 The integral over the angle of the twist kills the divergent holomorphic pole,
and renders the final result finite. The result is identical to the one 
obtained by covariant minimal subtraction (see  \cite{Giveon:2014hfa} for the detailed derivation of this statement).

\subsection{A Miniature}

Finally, we wish to assemble a miniature triptych. Firstly, we revisit
the path integral approach of section \ref{lattice} and apply it to
flat space. We T-dualize flat space, consider the infinite covering,
and find instead of the zero mode factor (\ref{zeromodefactor})
\begin{eqnarray}
Z_0^{\infty,flat} &=& 2 \pi N_{\infty} \int_0^{R} dr \partial_r (-\pi r^{-2}) \alpha e^{-r^{-2} \frac{\pi}{\tau_2} \alpha^2}
\nonumber \\
&=& 2 \pi N_{\infty} \frac{\tau_2}{\alpha} e^{- \frac{\pi}{ R^2 \tau_2  }\alpha^2} \, ,
\end{eqnarray}
where we have introduced an infrared cut-off $R$ on the radial integral.
Thus, we find for the infinite cover of the T-dual of flat space the infrared regulated elliptic genus
\begin{equation}
Z^{\infty,flat}(R) = N_{\infty} \frac{\theta_1(\alpha,\tau)}{\eta^3} \frac{1}{2 \pi \alpha} e^{- \frac{\pi \alpha^2}{R^2 \tau_2} } \, .
\label{flatcover}
\end{equation}
For flat space then, we find the same lattice sum (see equation (\ref{latticesum2})) as for the cigar elliptic genus, with the level $k$ replaced
by the infrared cut-off $R^2$. 

Our second panel, in section \ref{SQMhalf}, makes it manifest that
we have implicitly used the same boundary conditions for bosons and
fermions, since we considered a single measure, a hard infrared
cut-off $R$, and no delta-function insertion. Hence we find the anti-holomorphic $\bar{\tau}$ dependence in our result
(\ref{flatcover}).
Furthermore, our discussion in this section  agrees with
the fact that if we take the limit $R \rightarrow \infty$ term by
term, neglecting the exponential factor in (\ref{flatcover}), then we find a divergent result. 
Indeed, the lattice sum will be divergent. 

Finally, we note that (at $R=\infty$) the genus can be regulated  in the manner of the Weierstrass
$\zeta$-function (which is the regulated lattice sum of $1/\alpha$). If we take that ad hoc route, the result can
be made holomorphic and non-modular, and equal to only the first term in (\ref{flatfinal}), using the formula
\begin{equation}
\zeta(\alpha,\tau)-G_2(\tau) \alpha 
%= \frac{1}{\alpha}+\sum_{(n,m) \neq (0,0)}
%(\frac{1}{\alpha-(n-m \tau)} + \frac{1}{n-m \tau}) 
=
 \frac{\partial_\alpha \theta_{1}(\alpha,\tau)}{\theta_1(\alpha,\tau)} \, ,
\end{equation} 
where $G_2$ is the second Eisenstein series (and multiplying in the prefactor $\theta_1(\alpha,\tau)/\eta^3)$). %\footnote{The overall sign of the elliptic genus is conventional.}
%(The formula
%can be derived using the fact that the $\zeta$-function is the logarithmic derivative of the Weierstrass $\sigma$-function.)
On the other hand, if we infrared regulate with a radial cut-off as in
(\ref{flatcover}), or using the cigar model in the large level limit,
we obtain the modular covariant, non-holomorphic result
(\ref{flatfinal}) which equals the exponentially regulated Eisenstein
series as proven in \cite{Giveon:2014hfa,Eguchi:2014vaa}. This
final  miniature illustrates how our conceptual triptych folds together
seamlessly.

\section{Conclusion}

Our aim in this paper was to further explain conceptual features of
completed mock modular non-compact elliptic genera \cite{Troost:2010ud} with elementary means. Using the
supersymmetric cigar conformal field theory as an example, we provided
a simple path integral derivation of the lattice sum formula \cite{Eguchi:2014vaa} for
the completed mock modular form. We derived the elliptic genus from the non-linear sigma-model\footnote{Other derivations
are based on the coset conformal field theory or the gauged linear sigma-model \cite{Murthy:2013mya,Ashok:2013pya}
descriptions.}. We also laid bare the unresolvable tension
between right-moving supersymmetry and modular covariance in defining
the weighted trace with an infrared regulator, and we analyzed the
quirks of the identification of the large level limit of the cigar
model \cite{Giveon:2014hfa} with a flat space conformal field theory.

We believe these conceptual pointers provide a looking glass with
which to revisit higher dimensional elliptic genera, including the
$K3$, the ALE  \cite{Harvey:2014nha} and the higher dimensional
linear dilaton space genera \cite{Ashok:2013zka}. The ubiquitous
possibility to factor the appropriate powers of $\theta_1/\eta^3$ bodes well for this enterprise.
For four-dimensional examples, for instance, we expect the doubling of the number of right-moving zero modes 
to be correlated to an elliptic Weierstrass
$\wp$ factor in the result, et cetera. It will be interesting to study these generalizations.

\section*{Acknowledgments}
Many thanks to Sujay Ashok, Costas Bachas, Amit Giveon, Dan Israel, Sunny Itzhaki, Sameer Murthy, 
Boris Pioline,  Giuseppe Policastro, Ashoke Sen and Yuji Sugawara for useful discussions over the years, including
at the workshop on Mock Modular Forms and Physics in Chennai, India in 2014. I
acknowledge  support from the grant ANR-13-BS05-0001.

\end{document}